# Novel Superconducting SrSnP with Strong Sn-P Antibonding Interaction: Is the Sn Atom Single or Mixed Valent?

*Xin Gui,[a] Zuzanna Sobczak,[b] Tay-Rong Chang,[c] Xitong Xu,[d] Angus Huang,[e] Shuang Jia,[df] Horng-Tay Jeng,[eg*] Tomasz Klimczuk,[c] Weiwei Xie [a*]*

[a] Department of Chemistry, Louisiana State University, Baton Rouge, LA, USA 70803
[b] Faculty of Applied Physics and Mathematics, Gdansk University of Technology, Narutowicza 11/12, Gdansk, Poland 80–233
[c] Department of Physics, National Cheng Kung University, Tainan, Taiwan 30013
[d] International Center for Quantum Materials, School of Physics, Peking University, China 100871
[e] Department of Physics, National Tsing Hua University, Hsinchu, Taiwan 30013
[f] Collaborative Innovation Center of Quantum Matter, Beijing, China 100871
[g] Physics Division, National Center for Theoretical Sciences, Hsinchu, Taiwan 30013

## *ABSTRACT*

The large single crystals of SrSnP were prepared using Sn self-flux method. The superconductivity in the tetragonal SrSnP is observed with the critical temperature of ~2.3 K. The results of a crystallographic analysis, superconducting characterization, and theoretical assessment of tetragonal SrSnP are presented. The SrSnP crystallizes in the CaGaN structure type with space group *P*4/*nmm* (S.G.129, Pearson symbol *tP*6) according to the single crystal X-ray diffraction characterization. A combination of magnetic susceptibility, resistivity, and heat capacity measurements confirms the bulk superconductivity with $T_c$ = 2.3(1) K in SrSnP. According to the X-ray photoelectron spectroscopy (XPS) measurement, the assignments of $Sr^{2+}$ and $P^{3-}$ are consistent with the chemical valence electron balance principles. Moreover, it is highly likely that Sn atom has only one unusual oxidation state. First-principles calculations indicate the bands around Fermi level are hybridized among Sr-*d*, Sn-*p*, and P-*p* orbitals. The strong Sn-P and Sr-P interactions pose as keys to stabilize the crystallographic structure and induce the superconductivity, respectively. The physics-based electronic and phononic calculations are consistent with the molecular viewpoint. After including the spin-orbit coupling (SOC) into the calculation, the band degeneracies at Γ-point in the first Brillouin zone (BZ) split into two bands, which yield to the van Hove singularities around Fermi level.

## Introduction

The discovery of new materials that own novel properties, especially superconductivity, is a long-standing topic for solid-state material scientists.[1] There are various approaches to new superconductors, one of which is to assume that superconductivity appears in the structural families.[2] Many superconductors, including high-$T_c$ cuprates[3], boride carbides[4,5], and iron-based pnictides[6,7], crystalize in the layered structure, which is widely accepted to be one of the most critical factors to induce the superconductivity. The discovery of these new structural motifs hosting the superconductivity often leads to the rapid development of new superconductors. Both perovskites and the Fe-based superconductor structure types are good examples.[8,9] On the other hand, suppressing the charge disproportionation is another strategy to find new superconductors.[10] For example, the Mott insulator, $BaBiO_3$, can be chemically formulated as $Ba[Bi^{3+}]_{0.5}[Bi^{5+}]_{0.5}O_3$.[11] After doping K on Ba or Pb on Bi sites, the mixed valence state will vanish; meanwhile, the superconductivity will appear. In 2014, Hosono's group investigated the superconductivity in SnAs and focused on the valence state of Sn in SnAs to prove that Sn has the single valence state, likely $Sn^{3+}(5s^1)$, rather than a mixed valence state like $Sn^{4+}(5s^0)$ and $Sn^{2+}(5s^2)$.[12] Later in 2017, Tokura's group studied the superconducting SnP under high pressure and emphasized the critical role of the nominal valence $Sn^{3+}(5s^1)$ in stabilizing the phase and inducing superconductivity.[13,14]

Considering these anomalous oxidation states in Sn-based superconductors, we focused on ternary equiatomic SrSnP with an even shorter Sn-P bonding distance (~2.46 Å) compared to ~2.642 Å Sn-P in SnP at 2.2GPa.[15,13] A previously reported antiferromagnetic EuSnP has the isotype structure to SrSnP.[16] The crystal structure can be described as built up from tin atoms that form nets of puckered four-membered rings by means of long Sn-Sn bonds (~3.26 Å). Assuming the oxidation states of Sr to be +2 and P to be -3, each Sn will be left with only three valence electrons for form four equivalent Sn-Sn bonds in the puckered net, if all bonds are assumed to be normal two-center two-electron (2c-2e) bonds. The observation of four long Sn-Sn bonds then implies some sort of fractional bonding in which the bonds are weaker than normal 2c-2e bonds.[17,18] This will lead to an interesting question regarding the Sn valence: does the Sn atom have a single valence as we proposed above or mixed valences, like Sn (0) $(5s^25p^2)$ and Sn (+2) $(5s^2)$? Therefore, can the unusual single valent Sn be related to some unique physical properties, such as superconductivity?

With these structural, valent, and superconducting features in mind, we report herein a thorough structural characterization, investigation of the superconducting properties, and theoretical electronic and phononic structures of SrSnP, with an emphasis on the Sn valence and interplay between superconductivity and chemical bonding. We discovered the superconductivity in SrSnP with $T_c$ = 2.3(1) K for the first time. What's more, Sn has only one valent state, not mixed valent states like $Sn^0(5s^25p^2)$ and $Sn^{2+}(5s^2)$. A combination of chemistry and physics interpretations confirms both Sr-P and Sn-P bonding interactions are critical for the structural stability and superconductivity.

*Experimental Section*

**Single Crystal Growth of SrSnP:** The single crystals of SrSnP were grown from Sn-flux. Elemental strontium (99%, granules, Beantown Chemical), red phosphorus (99%, ~ 100 meshes, Beantown Chemical), and tin granules (99.5%, Alfa Aesar) were put into an alumina crucible with the molar ratio of 1:1:10. The crucible was subsequently sealed into an evacuated ($10^{-5}$ torr) quartz tube. A treatment was performed at a rate of 30 °C/h to 600 °C and held for 12 hours to avoid the explosion of red phosphorus. After that, the mixture was heated up to 1050 °C and annealing for 24 hours followed by slowly cooling to 600 ºC at a rate of 3 °C per hour. Excess Sn flux was removed by centrifuging the samples. Single crystals of SrSnP with a size of ~0.2×2×2 $mm^3$ tetragonal shape were obtained as shown in the insert of FIG. 1. All operations were carried out in the Ar-protected glovebox because SrSnP is extremely unstable toward decomposition in air or moisture.

**Phase Identification:** A Rigaku MiniFlex 600 powder X-ray diffractometer (XRD) equipped with Cu $K_\alpha$ radiation ($\lambda$=1.5406 Å, Ge monochromator) was used to examine the phase information. Due to the fact that the SrSnP single crystal is extremely air- and moisture- sensitive, we were only able to perform a short scan which the Bragg angle ranged from 10° to 70° in a step of 0.010° at a rate of 6.0°/min. The powder XRD pattern shown in FIG. 1 (Main Panel) was matched with the calculated pattern generated from the single crystal X-ray data and smoothened using Savitzky-Golay smoothing filters.[19]

**Structure Determination:** Multiple pieces of single crystals (~20×40×40 $\mu m^3$) were picked up to perform the routine structural examination. The structure was determined using a Bruker Apex II diffractometer equipped with Mo radiation ($\lambda_{K\alpha}$= 0.71073 Å) at low temperatures (215 K and 100 K). Glycerol and liquid nitrogen gas were used to protect the sample which was mounted on a Kapton loop. Seven different angles were chosen to take the measurement with an exposure time of 15 seconds per frame and the scanning 2θ width of 0.5°. The direct methods and full-matrix least-squares on $F^2$ models with SHELXTL package were employed to solve the crystal structure.[20] Data acquisition was made *via* Bruker SMART software with the corrections on Lorentz and polarization effect.[21] Numerical absorption corrections using face-index model were approached by XPREP.[22]

**Scanning Electron Microscopy (SEM):** The determination of the chemical composition and stoichiometry were performed using a high-vacuum scanning electron microscope (JSM-6610 LV) and Energy-Dispersive Spectroscopy (EDS). Samples were mounted on the carbon tape in the argon protected glovebox and quickly moved into the SEM chamber, and then evacuated immediately. Multiple points and areas were taken to test on the same single crystal of SrSnP with an accelerating voltage of 15 kV and the collecting time of 100 seconds for each point/area via TEAM EDAX software.

**Superconducting Properties Measurements:** The Superconducting Quantum Interference Device (SQUID) (Quantum Design, Inc.) magnetometer is used to measure the temperature-dependent susceptibility under different magnetic fields with the temperature range of 1.8-10 K. The susceptibility is defined as $\chi$ = M/H. Here, M is the magnetization in units of emu, and H is the magnetic field applied in Oe. Both field cooling (FC) and zero field cooling (ZFC) methods were used to test susceptibility. Resistivity and heat capacity measurements were performed using a quantum design physical property measurement system (PPMS) between 1.8K and 300K with and without applied fields. Heat capacity was measured using a standard relaxation method.

**X-ray Photoelectron Spectroscopy (XPS):** A Kratos AXIS 165 XPS/AES equipped with standard Mg/Al and high-performance Al monochromatic source was used to characterize the chemical states of elements on the surface of SrSnP in an evacuated ($10^{-9}$ torr) chamber at room temperature. To keep the surface of our sample clean, an Ar ion gun was utilized to remove the possible oxidized layer of SrSnP.

**Electronic Structure Calculations**

**Tight-Binding, Linear Muffin-Tin Orbital-Atomic Spheres Approximation (TB-LMTO-ASA):** Calculations of Crystal Orbital Hamiltonian Population (-COHP) curves were performed by Tight-Binding, Linear Muffin-Tin Orbital-Atomic Spheres Approximation (TB-LMTO-ASA) using the Stuttgart code.[23–25] The convergence criterion was set as 0.05 meV and a mesh of 64 $k$ points was used to generate all integrated values.[26] In the ASA method, space is filled with overlapping Wigner-Seitz (WS) spheres. The symmetry of the potential is treated as spherical in each WS sphere with a combined correction on the overlapping part. The WS radii are: 2.081 Å for Sr; 1.486 Å for Sn; and 1.367 Å for P. Empty spheres are required for the calculation, and the overlap of WS spheres is limited to no larger than 16%. The basis set for the calculations included Sr 5$s$, 4$d$; Sn 5$s$, 5$p$; and P 3$s$, 3$p$ wavefunctions.

**Vienna Ab initio Simulation Package (VASP):** The electronic structure was computed based on projector augmented-wave (PAW) pseudopotentials that were adopted with the Perdew-Burke-Ernzerhof generalized gradient approximation (PBE-GGA) implemented in VASP.[27] The experimental lattice parameters were used to perform the calculations. The energy cutoff was 400 eV. The 11×11×6 Monkhorst-Pack $k$-points mesh was used in calculation. The spin-orbit coupling effects are included.

**Quantum Espresso:** The electron-phonon coupling was completed using Quantum Espresso code[28,29] based on Density Functional Perturbation Theory (DFPT)[30,31], which employs norm-conserving pseudopotentials. Reciprocal space integrations were completed over a 24×24×12 $k$-point mesh, 4×4×4 $q$-point mesh with the linear tetrahedron method. The cut-off energies were set at 40 Ry (400 Ry) for wave functions (charge densities). With these settings, the calculated total energy converged to less than 0.1 Ry per atom. The electron-phonon coupling strength $\lambda_{eq}$ was calculated using $\lambda_{qv} = \frac{1}{\pi N_F} \frac{\Pi''_{qv}}{\omega^2_{qv}}$, where $N_F$ is the density of states (DOS) at the Fermi level, and $\omega_{qv}$ is the phonon frequency of mode υ at wave vector $q$. The critical temperature $T_c$ can be estimated by McMillan formula[32], $T_c = \frac{\omega_{ln}}{1.20} exp\left[-\frac{1.04(1+\lambda)}{\lambda-\mu^*(1+0.62\lambda)}\right]$ where $\lambda = \sum_{qv} \lambda_{qv}$ , $\omega_{ln} = exp\left[\frac{2}{\lambda} \int d\omega \frac{\ln \omega}{\omega} \alpha^2 F(\omega)\right]$, and $\alpha^2 F(\omega) = \frac{1}{2} \int d\omega \lambda_{qv} \omega_{qv} \delta(\omega - \omega_{qv})$ with μ*=0.1.

## Results and Discussion

**Phase, Crystal Structure and Chemical Composition Determination:** According to the previous phase study in Sr-Sn-P system, three different ternary phases included orthorhombic $Sr_3Sn_2P_4$ and $Sr_5Sn_2P_6$[15,33], and tetragonal SrSnP were reported. Our synthetic approach did not yield anything other than crystalline tetragonal SrSnP phase. The powder X-ray diffraction pattern of SrSnP was shown in FIG.1. It can be found that a pure SrSnP phase was obtained with slight Sn-flux impurity. The broad peaks located between 25-30° are due to the decomposition of SrSnP in the air. The result matches with the calculated pattern generated from our single crystal X-ray diffraction data and reference as well.[15]

To obtain further insights into the structural features of SrSnP, single crystals were investigated to extract atomic distributions and coordination environments. The results of single crystal diffraction including atomic positions, site occupancies, and isotropic thermal displacements are summarized in Tables S1 and S2. Corresponding anisotropic displacement parameters are summarized in Table S3. The result is consistent with the previous report. The lattice parameters are slightly smaller due to lower testing temperature (100K). Both vacant and mixed models were compared using Hamilton test[34] to confirm the full occupancy on each site. The chemical compositions of $Sr_{0.90(9)}Sn_{1.00(8)}P_{1.1(2)}$ examined by EDS are listed in Table S4. SrSnP crystallizes in the CaGaN structure type with the tetragonal space group *P*4/*nmm* which is the same as EuSnP.[16] Each Sn atom in the puckered sheet is surrounded by a distorted tetrahedron consisting of four Sn atoms with the Sn-Sn distance ~3.21Å. The connection to the upper and lower layers of the Sn layer is produced by an axial connection to the P atom at ~2.46 Å and an opposite axial connection to Sr at ~3.45 Å (FIG. 2 (A)). On adjacent Sn atoms the arrangement of axial Sr and P atoms is opposing, forming Sn-Sr-P-Sn stacking (FIG. 2 (B)). There are two layers comprised of Sr and P in a square array between every sheet of Sn atoms.

**SrSnP as an Elongation of SnP:** The structure of SrSnP can be regarded as the elongation after the insertion of Sr atoms into the cubic SnP frame followed by the Sr-induced bond inversion, as described in FIG. 3. To make the figure clear, we only show one quarter of the structure of cubic SnP. The binary phase SnP transforms to a tetragonal structure from the cubic NaCl-type structure by increasing pressure due to the tendency of insertion of Sr atoms. The structure of SrSnP can be considered as a derivative from the tetragonal SnP, both of which exhibit close Sn-P bonding interactions. By adding Sr to cubic SnP, half of the Sn-P bonds break and form the Sn-P hetero-dimers with a shorter Sn-P distance. Based on the fact that there is only one crystallographic atomic site for each atom, we will only describe the change on Sr1 atom surrounded the pyramid of P5@P1-4. Since the Sr has less electronegativity than Sn and P is more electronegative than Sn, it is likely to form Sr-P close bond rather than Sr-Sn close bond. Therefore, as shown in FIG. 3, Sn1-P6 has to be inversed along *c*-axis and build up a stronger Sr1-P6 bond. To prove this hypothesis, the total energy calculation of two different SrSnP models was performed. According to these relative total energies, the Model (A) for atomic interactions has a significantly lower total energy than Model (B), which has no Sn-P bond inversion. The theoretical assessment indicates good agreement with the experimental crystallographic results of SrSnP.

**Superconductivity in SrSnP:** FIG. 4 presents magnetic properties of SrSnP. Zero field cooling (ZFC) and field cooling (FC) temperature dependence of low field (H = 10 Oe) volume magnetic susceptibility is presented in panel (A). There are two clear superconducting transitions visible: at 3.2 K and 2.3 K that originate from Sn impurity and SrSnP, respectively. A small difference between the ZFC and FC curves suggests weak vortex pinning, that is typical for single crystal samples. Figure 4 (B) presents volume magnetization vs. applied magnetic field $M_V(H)$ measured in the superconducting state (T < $T_c$). At the low field $M_V(H)$ is almost symmetrical and suggests that SrSnP is a type-I superconductor.[35,36] To further characterize the superconductivity, FIG. 5 shows the temperature-dependent resistivity curve of SrSnP. The resistivity curve from 3 to 100 K indicates the metallic behavior of SrSnP without a phase transition. At 2.3 K, the resistivity of SrSnP undergoes a sudden drop to zero, which is a characteristic of superconductivity. To prove that the observed superconductivity is a bulk effect, the superconducting properties were characterized further through specific heat ($C_p$) measurements. Main panel of FIG 6 (A) shows $C_{el.}/T$ versus T in the superconducting transition region, where $C_{el}$ is an electronic specific heat estimated by subtraction of the lattice contribution ($C_{latt.}$) from the experimental data - $C_p$. Large anomaly that is visible in the specific heat data confirms bulk superconductivity of SrSnP. An equal area construction, denoted by solid lines, performed in order to determine superconducting transition temperature gave a value of $T_c$ = 2.3 K. This temperature is in agreement with results obtained from magnetic susceptibility and resistivity measurements. FIG 6 (B) presents the results of measurement of $C_p/T$ *vs* $T^2$ for the data obtained under 0 T and 50 mT magnetic field. The normal state was fitted to the equation $C_p/T = \gamma + \beta T^2$, represented by a solid line, where γ is an electronic contribution and β is a phonon contribution to the specific heat. The estimated Sommerfeld coefficient is equal to γ = 2.77(7) mJ mol$^{-1}$ K$^{-2}$ and β = 0.39(1) mJ mol$^{-1}$ K$^{-4}$, which is connected to the Debye temperature through the relation:

$$\Theta_D = (\frac{12\pi^4}{5\beta_3}nR)^{1/3}$$

where R = 8.31 J mol$^{-1}$ K$^{-1}$ and n = 3 for SrSnP. The Debye temperature calculated from this formula is $\Theta_D$ = 246(1) K. Knowing this value, allows calculation of the electron-phonon coupling constant from the inverted McMillan formula:

$$\lambda_{ep} = \frac{1.04 + \mu * \ln(\frac{\theta_D}{1.45T_C})}{(1 - 0.62\mu^*)\ln\left(\frac{\theta_D}{1.45T_C}\right) - 1.04}$$

Estimated value of $\lambda_{ep}$ = 0.55 suggests that SrSnP is a weak coupling superconductor. For this calculation the Coulomb repulsion constant was taken as μ* = 0.13. A weak coupling is also suggested by the normalized superconducting anomaly jump, that was calculated to be about $\Delta C/\gamma T_c$= 1.48, which is slightly above the expected value (1.43) for a weak-coupling BCS superconductors.

Having both the electron-phonon coupling constant $\lambda_{el-ph}$ and the Sommerfeld coefficient $\gamma$, the density of states at the Fermi energy can be calculated from:

$$N(E_F) = \frac{3\gamma}{\pi^2 k_B^2 (1+\lambda_{ep})}$$

The value for SrSnP is 0.86 states eV$^{-1}$ per f.u. FIG. 6 (C) presents the temperature dependence of $C_p$ divided by T$^3$. A maximum of $C_p/T^3$ occurs at about $T_{max} \approx 14$ K, and since $T_{max} = \Theta_E/5$, the Einstein temperature is estimated to be $\Theta_E \approx 70$ K.

**Single Valent or Mixed Valent of Sn?** With the question of the valence state of Sn in mind, we performed the X-ray Photoelectron Spectroscopy (XPS) experiment on SrSnP. FIG. 7 shows the XPS spectrum of Sn $3d_{5/2}$ core-level region. The asymmetric line shape of Sn $3d_{5/2}$ is due to the metallic properties.[37] To clarify the valence state of Sn, we assume Sn has mixed valences of Sn$^0$ and Sn$^{2+}$. Using this hypothesis, the fitting peaks of the Sn $3d_{5/2}$ core-level are 484.62 eV and 485.40 eV with the atomic ratio of 5.18:1 between Sn$^0$ and Sn$^{2+}$. However, according to FIG. S1, the shape of Sn$^{2+}$ peak we assumed is not reasonable due to the broad characteristic and the full width at half maximum (fwhm) of peaks are too large (1.43 & 1.59 eV), which confirms the single valent hypothesis of Sn. The peak position of Sn $3d_{5/2}$ (484.65 eV) is slightly higher than Sn metal (484.38 eV) which is also tested as standard (see FIG. S2) but lower than Sn-related oxides such as SnO (485.8 eV)[38]. The measurement indicates that Sn in SrSnP has only one valence state and the XPS result is consistent with that SrSnP has only one Sn site without any structural distortion. Is Sn in the +1 oxidation state? Usually, Sn$^0$ with 4 valence electrons would be four-bonded with other Sn atoms, and Sn$^{+1}$ will be left with only three valence electrons to form four Sn-Sn bonds with puckered nets, if all bonds are assumed to be normal two-center two-electron (2c-2e) bonds. However, the observation of four long Sn-Sn bonds implies partially fractional bonding in which the bonds are weaker than normal 2c-2e bonds. To further confirm this observation, we integrated the partial DOS of Sn, ~ 2.39 and 2.38 valence electrons per Sn without and with SOC, respectively.

**Electronic Structure of SrSnP:** To gain further insights into the electronic influences on the stability and superconducting properties of SrSnP, VASP calculations were carried out to evaluate and analyze the band structures. FIG. 8 (A) shows the first Brillouin zone (BZ) of a primitive tetragonal lattice with the high-symmetry $k$ points marked. The calculated bulk band structure of SrSnP, generated here with the inclusion of spin-orbit coupling (SOC), is shown in FIG. 8 (B). The band structure calculation indicates that the Sn $6p$ and P $3p$ orbitals dominate the conduction and valence bands around the Fermi level ($E_F$). The Fermi surface is predominantly comprised of two symmetric electron pockets along Z-A-Z and Γ-M-Γ and saddle points around Γ point, which are significant for the superconductivity. The "symmetric" band structure is usually originated from the band folding in the first Brillouin zone. Furthermore, the band structure of first BZ applies for the crystal momentum in the compound. To evaluate the real momentum in SrSnP more accurately, we performed the unfolding band structure calculation in the second Brillouin zone. With the Sn-Sr-P-Sn stacking along $c$-axis, the closed Sn-Sr-P-Sn chain is repeating along body-diagonal on the *ab*-plane shown in FIG. 8 (C). In the second BZ picture, electrons encounter Sn atoms in the crystal and form the Bloch waves. The electron interactions with P or Sr can be treated as a small perturbation to the Block wave. Thus, the second Brillouin zone and band structure are generated in FIG. 8 (D) and (E). The unfolding of the new BZ marked in a blue line shows the actual measured gap is at the M point. A big band gap appears at Γ point instead. To evaluate the SOC effects on the electronic structure of SrSnP, we also calculated the Density of States (DOS) shown in FIG. 9. Interestingly, SOC opens up the electronic gap, but it has no effect on $N_F$ (1.59 states/eV for both).

**Bonding Interactions in SrSnP:** To understand the atomic interactions in the SrSnP, the corresponding –COHP curves were calculated and presented in FIG. 10 (A). It shows that Sr-P bonding and Sn-P antibonding interactions govern the atomic interactions contribution around the Fermi level while Sn-Sn bonding interaction with Sn-Sn ~3.21Å is not as significant as Sr-P or Sn-P interactions. Moreover, the Sr-Sn interactions are barely found. This indicates the importance of Sr-P bonding interaction in stabilizing the structure and the possibility for Sn-P antibonding interaction to influence the physical properties most. Moreover, the Sn-P bonding interactions are dominated below the Fermi level along with Sr-P bonding interactions, which signifies that the crucial bonding interaction in SrSnP are Sn-P and Sr-P bonds. Corresponding to the previous structural study between SnP and SrSnP, it can be shown that the valence electron contribution from Sr atom overlaps with Sn-P σ-like antibonding part (FIG. 10 (B)). Such overlapping weakens the Sn-P bonding interaction and possibly causes the partially bond breakage of Sn-P which leads to the unusual oxidation state of Sn in SrSnP. This can be an analogy to Sn in SnP or SnAs[12-14] due to the strong electron transfer from Sn to P or As in the previous study. Therefore, the lower oxidation state delocalizes more valence electron on Sn atoms which may strengthen electron-phonon coupling and induce the superconductivity.

**Electron-Phonon Coupling in SrSnP:** To test whether this molecule-like hypothesis works for SrSnP in the context of the electron-phonon coupling expected from a physics-based picture, the systematic calculations for the electron-phonon coupling of SrSnP was calculated by use of the program Quantum Espresso. FIG. 11 (A) and (B) show the phonon spectrums of SrSnP with and without SOC. It can be clearly noted that the most contribution comes from Sn and SOC affects the phonon band at M point significantly, which also hosts the saddle points in the electronic band structure. In our calculations in FIG. 12, we find the total electron-phonon coupling strength $\lambda_{el-ph}$ to be 0.59 without SOC and 0.62 with SOC, which is consistent with the experimental evaluation. Using the McMillan formula, the superconducting transition temperatures $T_c$s are estimated to be 1.95 K SrSnP without SOC and 2.15 K for SrSnP with SOC. The theoretical prediction of $T_c$ with SOC matches with our experimental result $T_c = 2.3$ K very well, which indicates the SOC can help to improve the $T_c$.

**Spin-Orbit Coupling (SOC) on $T_c$:** How does SOC improve the $T_c$? Derived from the BCS theory, the superconducting critical temperature can be expressed as $kT_c = 1.13\ \hbar\omega\exp(-1/N(E_F)V)$, where V is a merit of the electron-phonon interaction, $N(E_F)$ is the DOS at Fermi level, and ω is a characteristic phonon frequency. The SOC usually affects the superconducting transition temperature by influencing $N(E_F)$ or ω. In SrSnP, since the SOC gap is near the high symmetry point, the impact of SOC on N ($E_F$) is quite small. Instead, the SOC affects ω and increase the $T_c$ in SrSnP.

## *Conclusion*

Herein, we successfully grew the single crystal of reported ternary compound SrSnP with a tetragonal space group of *P*4/*nmm*. Physical properties measurements confirm the existence of bulk superconductivity, which is consistent with the electronic and phononic calculation prediction. We used XPS measurements to examine the single valent state of Sn in SrSnP, which is critical for the superconductivity. From a chemistry viewpoint, the Sr-P bonding and Sn-P antibonding interactions are related to the anomalous oxidation state of Sn and superconductivity in SrSnP. The systematic electron-phonon coupling calculation from physics perspectives matches well with the molecule-based chemistry picture. The experimental and partial theoretical superconducting and thermodynamic parameters of SrSnP are summarized in Table 1. More interestingly, the strong spin-orbit coupling increases the critical temperature in SrSnP.

## *Supporting Information*

The Supporting Information is available free of charge on the ACS Publications website at DOI: xxxxx.

Crystallographic data, atomic coordinates, equivalent isotropic displacement parameters, anisotropic thermal displacement, SEM data, XPS spectra of SrSnP and standard Sn.

## *Acknowledgement*

X.G. and W.X. deeply appreciate the thoughtful suggestions from reviewers and useful discussion with Dr. Tai Kong (Princeton University), Dr. Cheng-Yi Huang and Dr. Hsin Lin (Institute of Physics, Academia Sinica), and technical support from Louisiana State University-Shared Institute Facility (SIF) for SEM-EDS and XPS. X.G. was supported by the State of Louisiana-Board of Regents Research Competitiveness Subprogram (RCS) under Contract Number LEQSF (2017-20)-RD-A-08. W.X. thanks the support through the Beckman Young Investigator (BYI) Program. The research in Poland was supported by the National Science Centre (grant UMO-2016/22/M/ST5/00435). X.X. and S.J. were supported by National Natural Science Foundation of China No. 11774007 and the Key Research Program of the Chinese Academy of Sciences (Grant No. XDPB08-1). T.R.C. was supported by the Ministry of Science and Technology and National Cheng Kung University, Taiwan. A.H. and H.T.J. was supported by the Ministry of Science and Technology, National Tsing Hua University, and Academia Sinica, Taiwan. A.H., T.R.C. and H.T.J. also thank NCHC, CINC-NTU and NCTS, Taiwan for technical support.

**Table 1.** Experimental and partial theoretical superconducting and thermodynamic parameters of SrSnP.

| Parameters | Experimental | Theoretical |
|---|---|---|
| $T_c$ (K) | 2.3 | 1.95 (with SOC) <br> 2.15 (without SOC) |
| $H_c$ (Oe) | 44(1) | \ |
| $\gamma$ (mJ mol$^{-1}$ K$^{-2}$) | 2.77(7) | \ |
| $\beta$ (mJ mol$^{-1}$ K$^{-4}$) | 0.39(1) | \ |
| $\Theta_D$ (K) | 246 (1) | \ |
| $\Delta C/\gamma T_c$ | 1.48 | \ |
| $\lambda_{el-ph}$ | 0.55 | 0.62 (with SOC) <br> 0.59 (without SOC) |

**FIG. 1. (Main Panel)** The powder X-ray diffraction pattern of SrSnP where the blue and red lines represent observed pattern and calculated patterns based on single crystal data, respectively. Trace Sn residual can be found. **(Insert)** The single crystal of SrSnP with dimension.

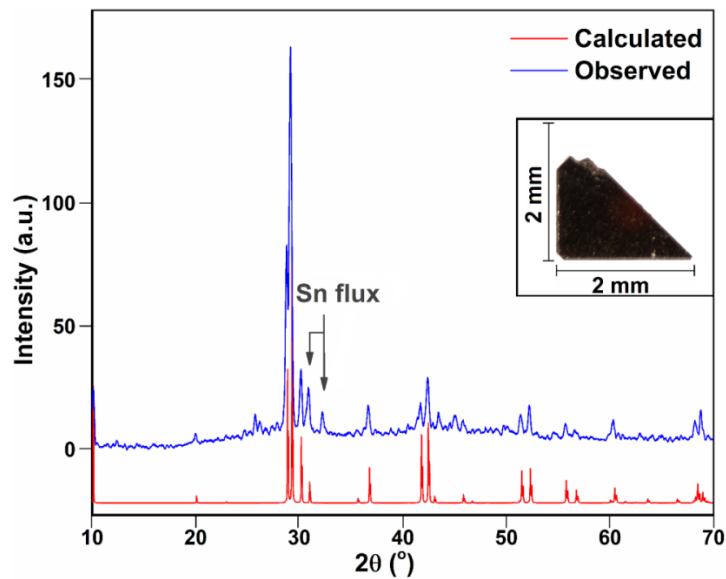

**FIG. 2. (A).** Crystal structure of SrSnP where the green, grey and red balls represent Sr, Sn and P atoms, respectively. **(B).** Sn-Sr-P-Sn stacking along *c*-axis labeling with bond lengths.

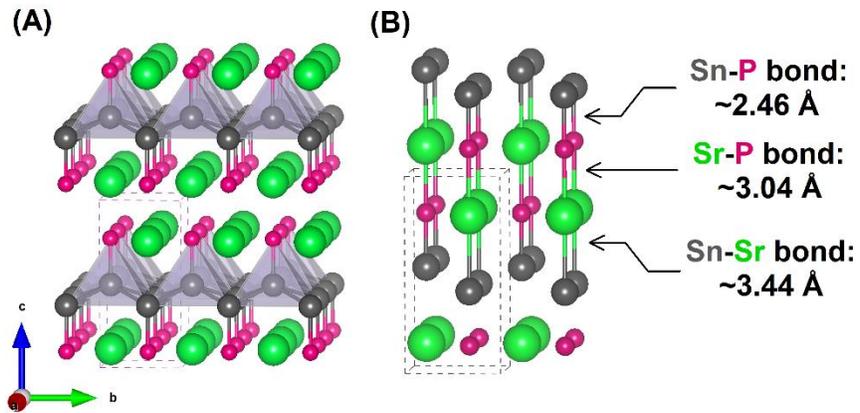

**FIG. 3.** The structure relationship between cubic SnP, tetragonal SnP and tetragonal SrSnP. Model **(A)** and **(B)** represent the experimental obtained SrSnP and the schematic figure for SrSnP without Sn-P bond inversion, respectively.

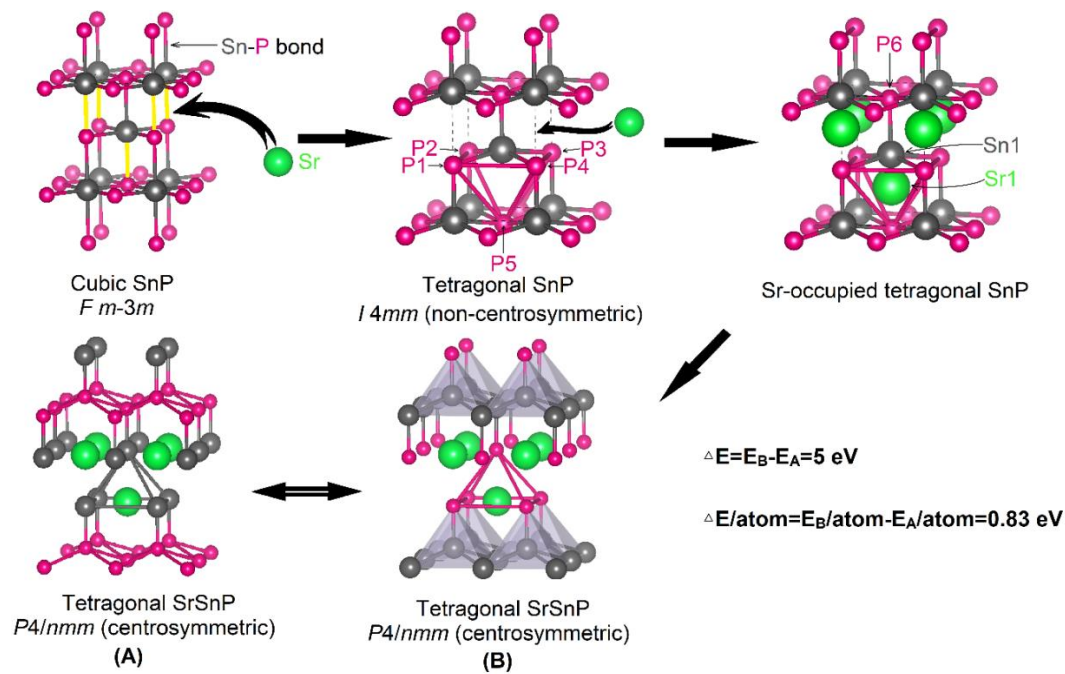

$\triangle E = E_B - E_A = 5$ eV

$\triangle E/\text{atom} = E_B/\text{atom} - E_A/\text{atom} = 0.83$ eV

**FIG. 4. (A).** Temperature (T) – dependent magnetic susceptibility ($4\pi\chi_v$) between 1.8-10 K performed with zero-field cooling (ZFC) and field-cooling (FC) modes. **(B).** Magnetization (M) vs. applied field (H) at multiple temperatures.

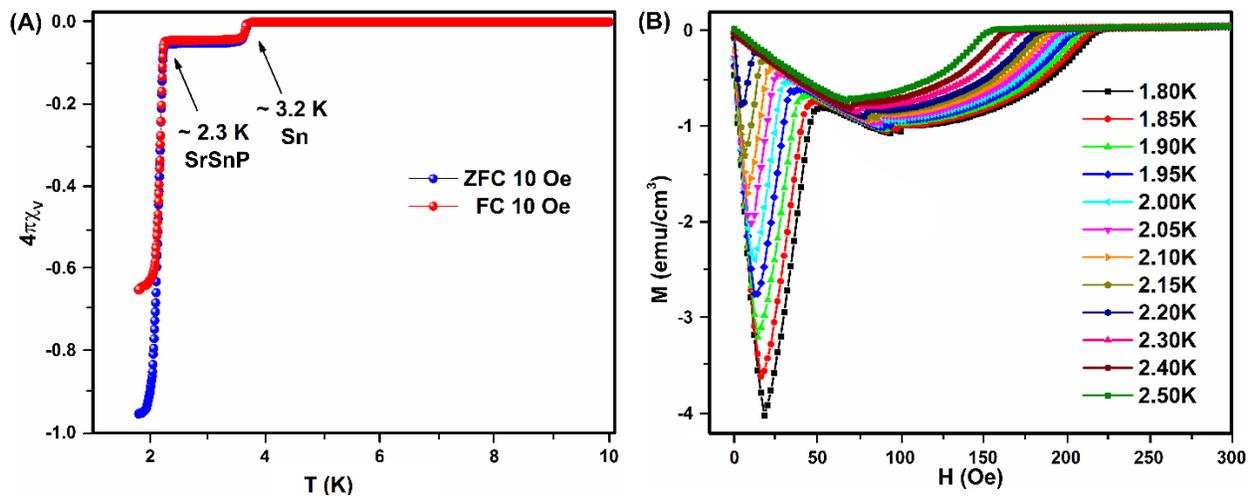

**FIG. 5. (Main Panel).** Temperature-dependent resistivity of SrSnP with the temperature ranging from 1.8 to 100 K. **(Insert).** Resistivity vs temperature at low-temperature region (1.8-3.0 K).

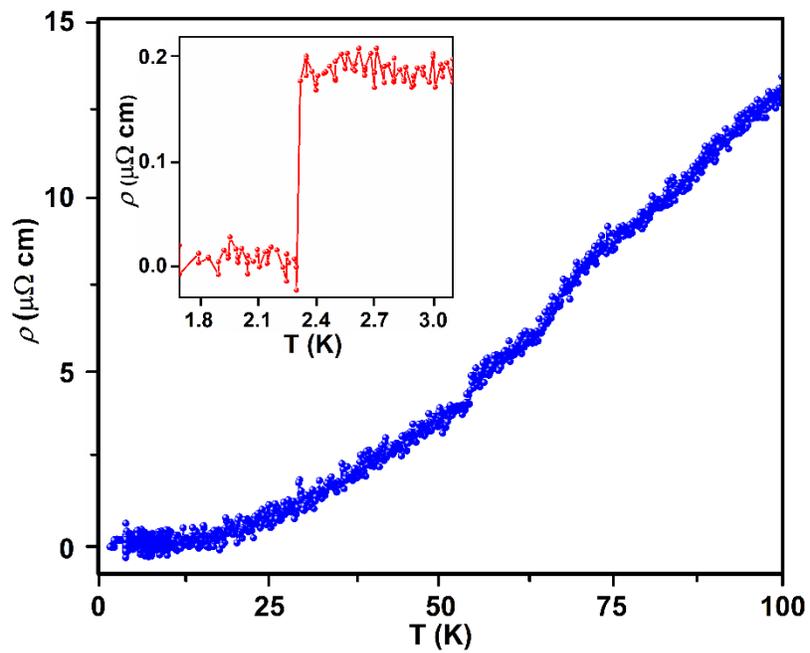

**FIG. 6. (A).** Temperature dependence of $C_{el}/T$ in superconducting transition region. **(B).** $C_p/T$ *vs* $T^2$ with/without the applied field of 50 mT. A straight solid line represents a fit $C_p/T = \gamma + \beta T^2$ **(C).** Temperature dependence of $C_p/T^3$.

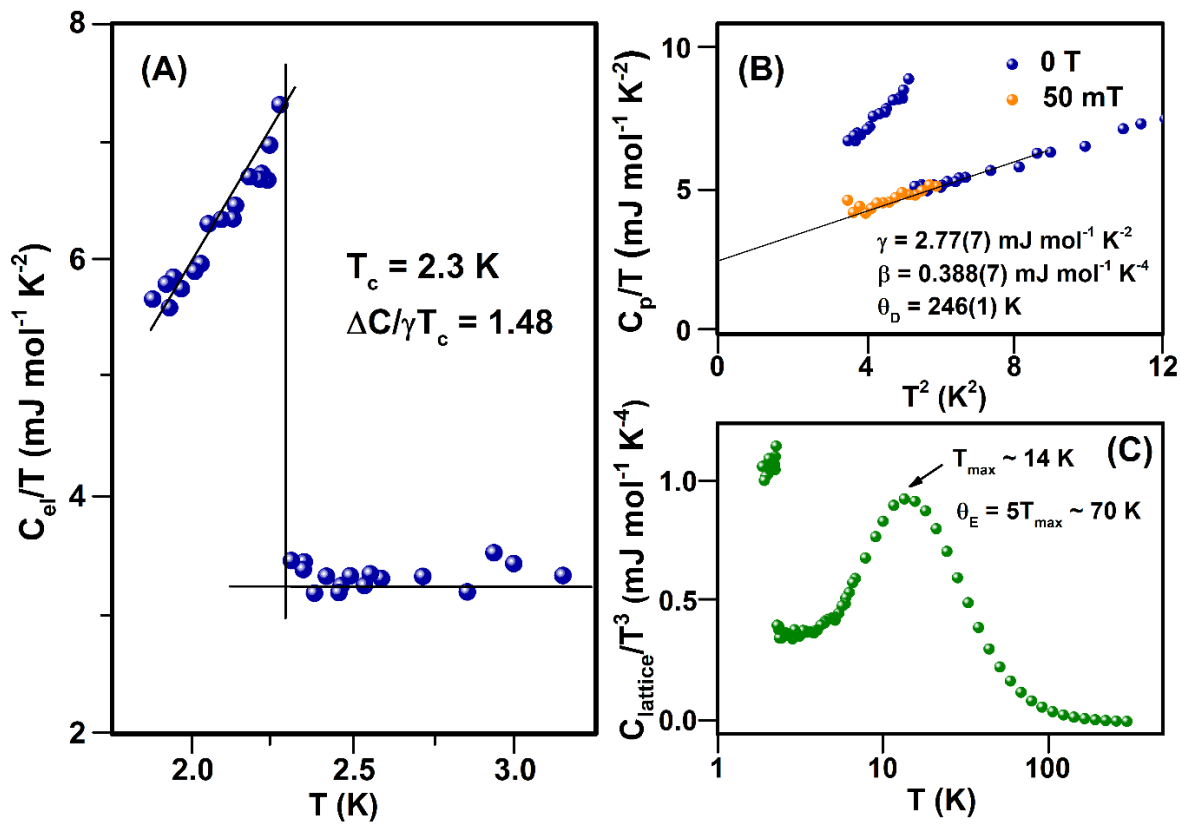

**FIG. 7.** XPS spectra of Sn *3d* core-level. The red and blue lines indicate experimental and fitting data, respectively. The *x*-axis represents binding energy (eV) and *y*-axis is the photonelectron intensity (counts per second (CPS)).

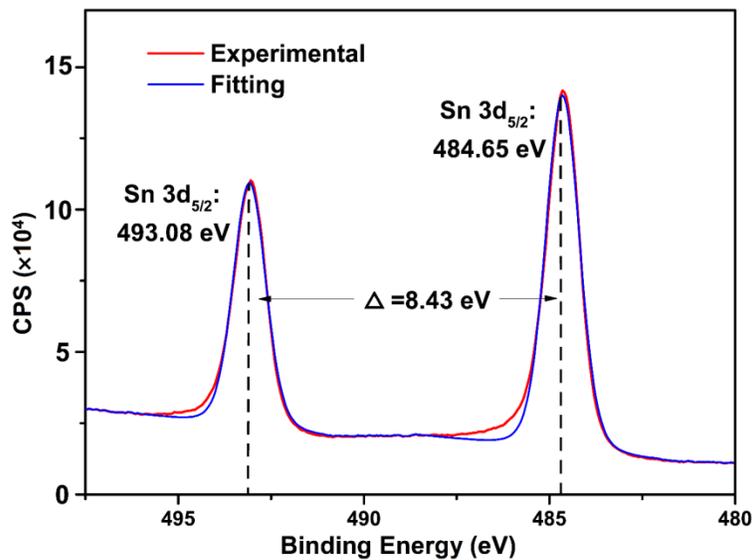

**FIG. 8.** **(A)**. The first Brillouin zone (BZ) of a primitive tetragonal lattice with high-symmetry *k* points. **(B)**. Band structure of SrSnP with SOC. **(C)**. Repeating Sn-Sr-P-Sn chain along the body diagonal of *ab*-plane. **(D)**. The second BZ of primitive tetragonal lattice. **(E)**. Band structure of SrSnP on the second BZ with SOC.

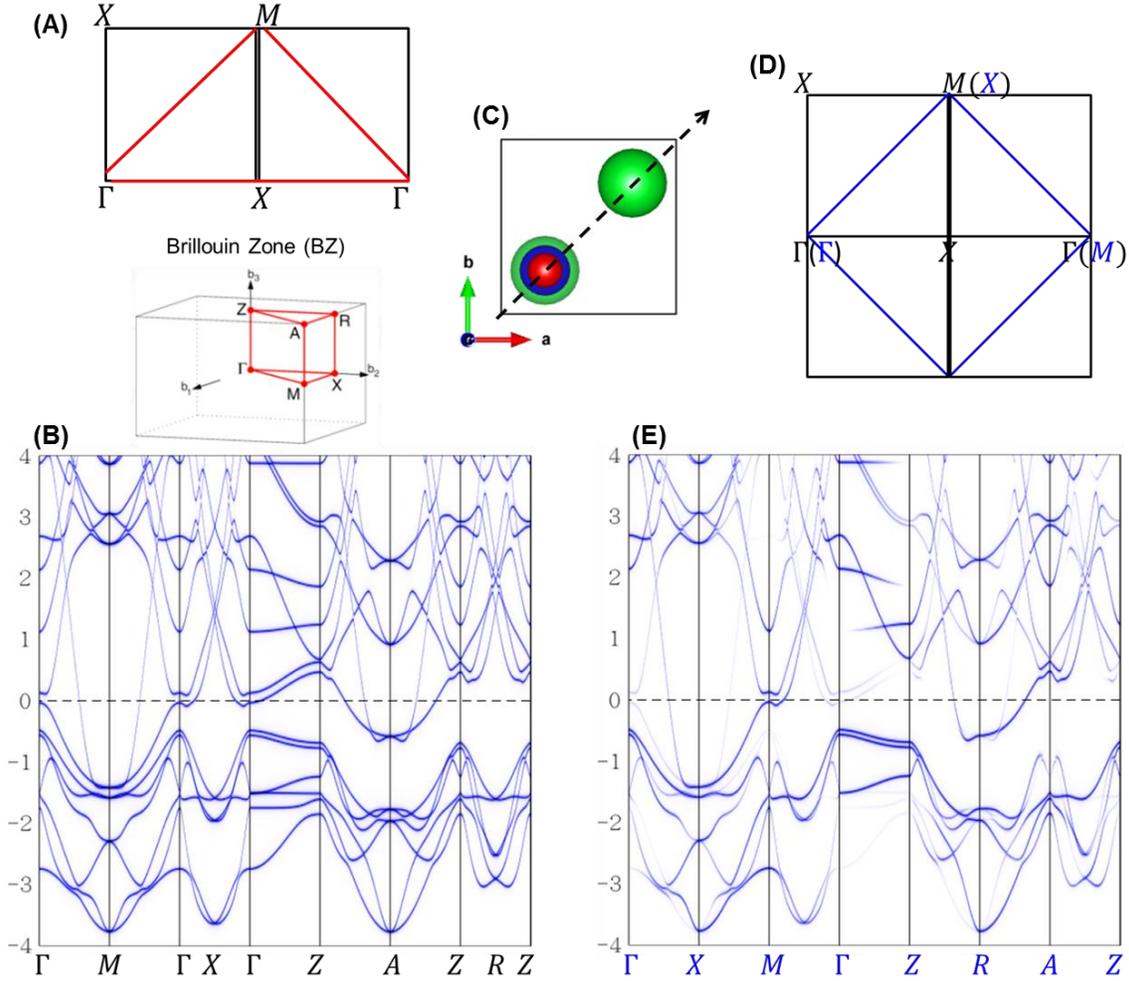

**FIG. 9.** Density of states (DOS) calculated with/without SOC.

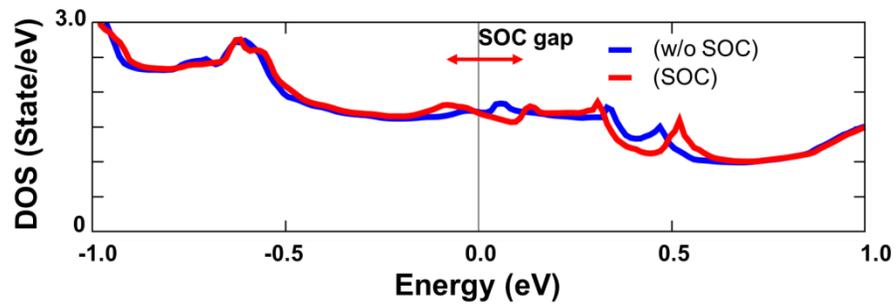

**FIG. 10. (A)**. Crystal Orbital Hamiltonian Population (-COHP) calculation for SrSnP. **(B)**. The electrons on $d_{x^2-y^2}$-orbital of Sr atoms are overlapping with Sn(P)-$p_x/p_y$ antibonding orbitals at Γ point near the Fermi level which may weaken the Sn-P bonding interaction to create more itinerant electrons of Sn. Higher nodal planes are omitted for orbitals.

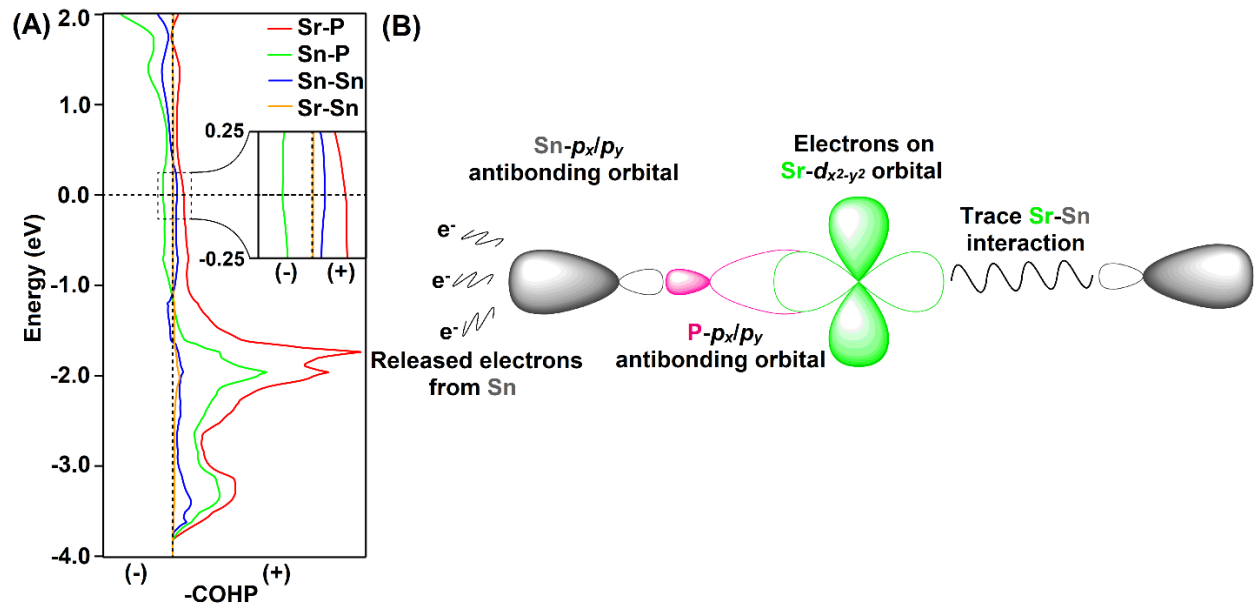

**FIG. 11. (A).** The phonon spectrums of SrSnP with and without SOC. **(B).** Phonon mode with consideration of SOC for different atoms where the blue, red and green lines represent Sn, Sr and P atoms, respectively.

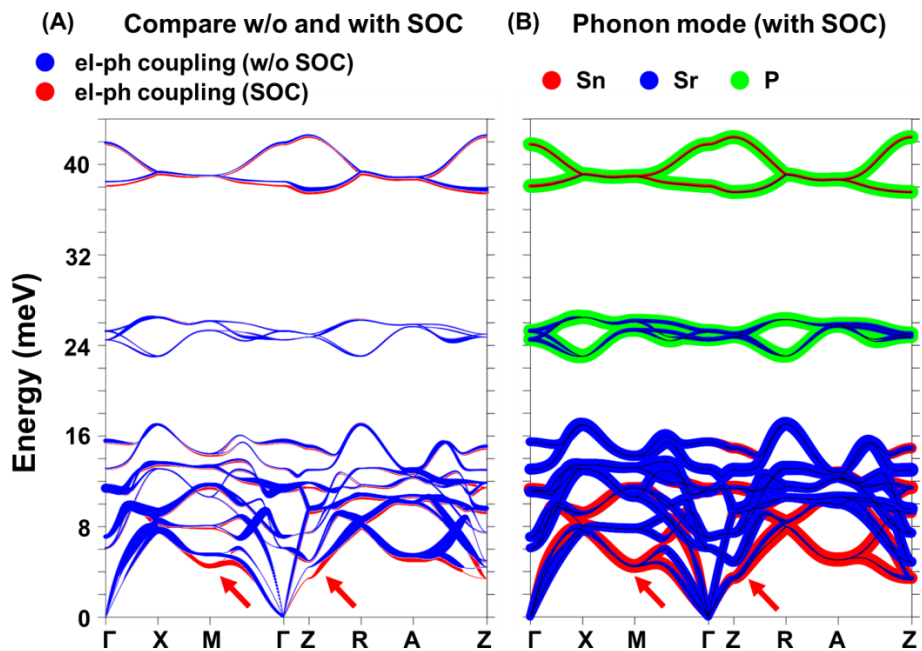

**FIG. 12.** Phonon band structure and Density of States (DOS) of SrSnP with/without SOC to generate the λ and predict the critical temperatures.

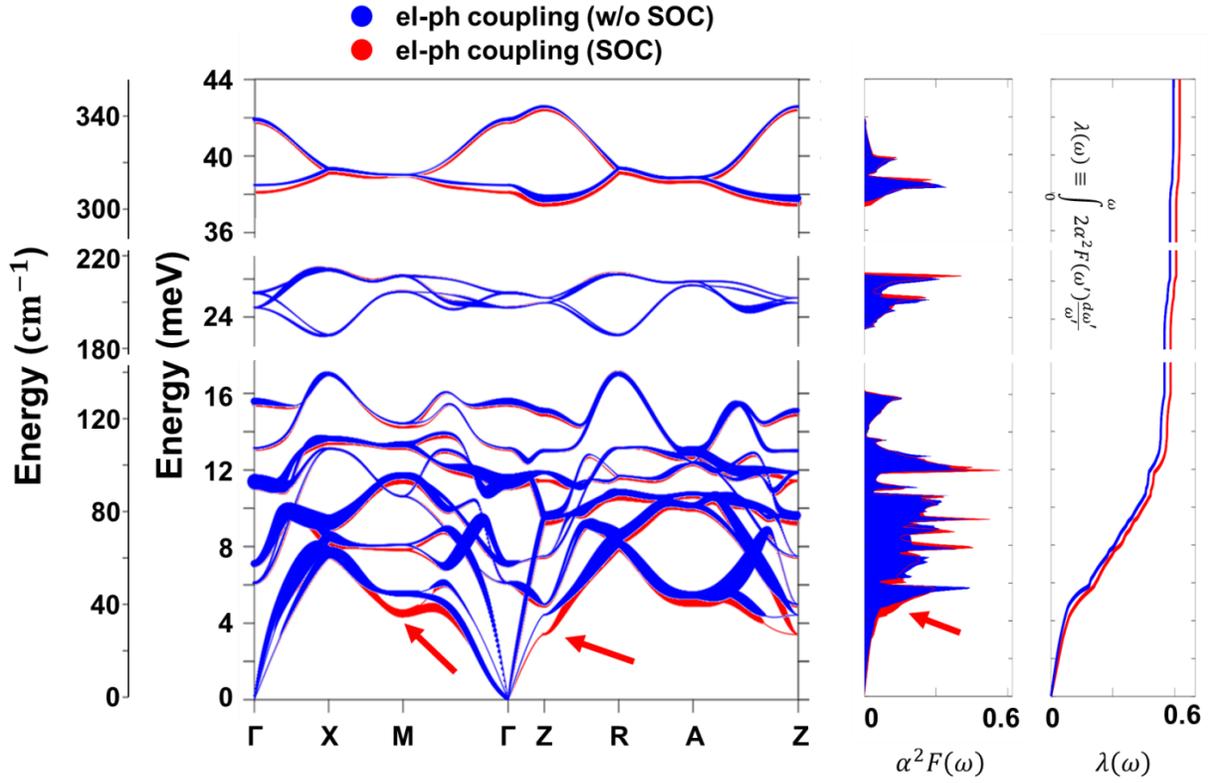

**For Table of Contents Only**

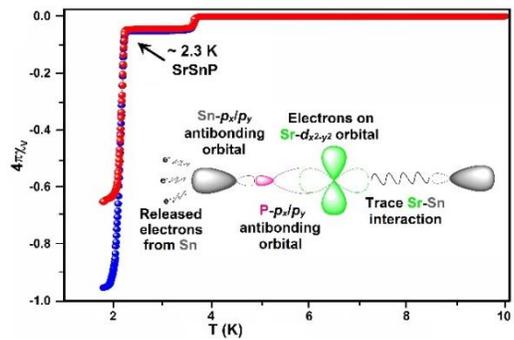